\documentclass[reprint,aip,cha,showpacs,amsmath,amssymb]{revtex4-1} % added by PXP Help
\usepackage{color}
\usepackage{graphicx}% Include figure files
\newcommand{\un}[1]{\ensuremath{\, \mathrm{#1}}}
\usepackage{fancyvrb}

\begin{document}

%Title of paper
\title[Special Issue for Chaos]{Experiments on autonomous Boolean networks}

\author{David P. Rosin}
\affiliation{Duke University, Department of Physics, Science Drive, Durham, North Carolina 27708 USA}
\affiliation{{Institut~f{\"u}r~Theoretische Physik{,}~Technische~Universit{\"a}t~Berlin{,}~Hardenbergstr{.}~36{,}~D-10623 Berlin{,}~Germany}}

\author{Damien Rontani}
\affiliation{Duke University, Department of Physics, Science Drive, Durham, North Carolina 27708 USA}

\author{Daniel J. Gauthier}
\affiliation{Duke University, Department of Physics, Science Drive, Durham, North Carolina 27708 USA}

\author{Eckehard Sch{\"o}ll}
\affiliation{{Institut~f{\"u}r~Theoretische Physik{,}~Technische~Universit{\"a}t~Berlin{,}~Hardenbergstr{.}~36{,}~D-10623 Berlin{,}~Germany}}

\date{\today}

\begin{abstract}
We realize autonomous Boolean networks by using logic gates in their autonomous mode of operation on a field-programmable gate array. This allows us to implement time-continuous systems with complex dynamical behaviors that can be conveniently interconnected into large-scale networks with flexible topologies that consist of time-delay links and a large number of nodes. We demonstrate how we realize networks with periodic, chaotic, and excitable dynamics and study their properties. Field-programmable gate arrays define a new experimental paradigm that holds great potential to test a large body of theoretical results on the dynamics of complex networks, which has been beyond reach of traditional experimental approaches.
\end{abstract}

\maketitle

% 05.45.-a = nonlinear Dynamical systems
% 64.60.aq = Networks in phase transitions
% 87.16.Yc = Regulatory networks in subcellular structure and processes
%07.50.Ek = electronic circuits
%89.75.-k = Complex systems

\begin{quotation}
Dynamical networks have attracted considerable attention because of their ubiquitous presence in numerous fields,\cite{STR01a,NEW10} such as biology (cellular and metabolic networks, food webs, neural networks)\cite{BAR04a,JEO00,MON06a,BUL09} and social sciences (mobile communication networks, scientific collaboration networks).\cite{ONN07,NEW01c} Insight into the dynamics of networks comes predominately from studies of mathematical models and observations of real-world networks as a test bed for theoretical results. There is also need to realize networks in the laboratory to test theoretical predictions in a controlled environment. But hitherto, the difficulty to connect a large number of dynamical nodes has restricted experiments to coupling topologies with at most 20 nodes.\cite{NIX12,TEM12} As a solution, computer algorithms have been used to manage the coupling between experimental dynamical systems.\cite{HAG12,TIN12} Here, we present an approach without computer-assisted coupling for the experimental realization of networks of potentially large sizes using a field-programmable gate array (FPGA)---an integrated circuit with millions of reconfigurable logic gates. Using its autonomous mode of operation, we implement continuous-time dynamical systems with periodic, chaotic, and excitable dynamics that can be coupled to arbitrary topologies and display collective phenomena such as synchronization.
\end{quotation}

\section{Introduction}

Logic gates on field-programmable gate arrays (FPGAs) can be assigned to arbitrary Boolean functions\cite{BRO08} and interconnected to form Boolean networks, which are used typically to model diverse biological processes, such as gene and metabolism regulation,\cite{JAC61,KAU03a,SAH10} cell-cycle dynamics,\cite{LI04b} neural interactions,\cite{BOR00b} and social networks.\cite{PAC00}

The Boolean states of the nodes in a Boolean network evolve in time according to logic functions.\cite{KAU93} Typically, the dynamical state of the network is updated either synchronously or asynchronously, where the Boolean states of the nodes are updated according to their logic functions simultaneously or successively with randomly chosen updating order, respectively.\cite{GRE05} These updating strategies for the network dynamics simplify the mathematical analysis\cite{POM09} and allow for exact numerical simulations but are, in some regard, unrealistic.

For example, gene and metabolism regulation networks are not updated by a discrete global clock in nature and should, therefore, be modeled continuously in time with dynamical updating of the the logic functions.\cite{MES97} To account for continuous temporal evolution, Boolean delay equations (BDEs)\cite{GHI85,GHI08} and ordinary differential equations\cite{MES96a} have been introduced and the networks are then referred to as autonomous Boolean networks (ABN). A node in an ABN updates its Boolean state whenever Boolean transitions are present at its inputs. Because of the finite response time of the node, intermediate output states have been numerically and experimentally observed. The consequence of the autonomous operation and the resulting non-ideal behavior of the gate 
is the existence of rich and complex dynamics, such as chaos\cite{ZHA09a,CAV10} and quasi-periodicity.\cite{GLA98}

FPGAs allow one to realize experimentally large ABNs to test theoretical predictions\cite{MES96a,MES97,GLA98} of models of Boolean networks. The experimental approach reveals dynamics that is not predicted in theoretical studies, which usually neglect non-ideal behaviors that may appear both in the biological and electrical systems, such as the sigmoidal activation functions of the gates, intrinsic parameter heterogeneity, and noise. Glass \emph{et al.\ }have already identified differences between the dynamics of an idealized model and the electronic implementation of a simple Boolean network.\cite{GLA05} 

Realizing large ABNs on an FPGA also has the advantage of fast dynamics, where the network nodes evolve with fast rise and fall times on the order of $300\un{ps}$. In contrast, a numerical simulation of an ABN requires multiple calculations for the continuous rises and falls for each logic gate in the network, which usually takes time on the order of milliseconds with current computer technology. With our approach, the network dynamics are, therefore, much faster generated than with the corresponding simulation.

In this article, we demonstrate the potential of FPGAs to realize experimentally large-scale complex networks with controllable node dynamics and arbitrary topology. The networks are meta-networks consisting of interconnected autonomous logic circuits---an electronic realization of ABNs---that represent dynamical nodes with various types of dynamics. We first introduce the FPGA and its autonomous mode of operation. Then, we introduce a Boolean phase oscillator and couple it to networks that display phase synchronization. We continue by introducing an ABN that displays chaotic dynamics. Finally, we realize ABNs with excitable dynamics that can be coupled to large networks and synchronize with zero-time lag in the presence of coupling time delays.

\section{Experimental networks on a chip}

In this section, we introduce the main working principles of FPGAs. We detail how to implement time-continuous dynamical  nodes (ABNs) and connect them with time delay links into meta-networks.

\subsection{Programmable Logic Gates}

An FPGA has up to 2 million re-assignable logic gates that generate high and low voltages $V_{H,L}$ corresponding to the Boolean states 0 and 1. They can execute any of the $2^n$ possible logic functions, where $n$ is the number of inputs whose value is typically four or six depending on the FPGA technology. For our experimental platform (Altera Cyclone IV EP4CE115F29C7N), $n=4$. Logic operations with more inputs can be realized by combining multiple logic gates.

The logic circuit---the logic gates and their interconnection---is specified using a hardware description language, such as Verilog or VHDL.\cite{BRO08} A compiler optimizes and converts the logic design so that it can be loaded on the FPGA, thereby specifying the Boolean operation of each logic element and the manner in which they are connected. These operations take as little as a few seconds depending on the complexity of the design. The flexibility, the speed, and the large number of available logic gates render the FPGA a promising platform for the realization of network experiments with large network sizes and complex topologies.

\subsection{Autonomous Mode of Operation}

The mode of operation of the logic gates has important implications for the dynamics of the logic circuit on the FPGA. In most applications of FPGAs, they are used in the synchronous operation with a clock period slow enough so that all logic gates can settle to their Boolean states between two consecutive clock cycles.\cite{BRO08} Then, the logic gates behave in a digital fashion consistent with the Boolean algebra. In the autonomous mode of operation, in contrast, the logic circuit displays an analog dynamical evolution governed by the logic gates' propagation delays, gate activation function, and low-pass filtering characteristics.\cite{ZHA09a} These properties vary between the logic gates on a chip because of manufacturing imperfections. Consequently, two autonomous logic circuits of identical layout that are realized on different regions on the FPGA can display somewhat non-identical dynamics.

\subsection{Design of Time Delay Links and Dynamical Nodes}
\label{sec:design_nodes_links}%this is important for cross reference of the link construction

To build dynamical networks, we identify a circuit design for the dynamical nodes and the network topology using the built-in switching fabric of the FPGA. Direct links can be realized with on-chip wires that have a delay of a few tens of picoseconds, which can often be neglected compared to the propagation delay of logic gates $\tau_\mathrm{LG}=(280\pm10)\un{ps}$ (numeric values for the FPGA used in our experiments). Links with substantially longer time delay can also be realized by exploiting the finite propagation time of logic gates. Specifically, time delay links are built by cascading an even number of $n_k$ inverter gates to achieve a time delay of $\tau_{n_k}=n_k\tau_\mathrm{LG}$. A delay line built from an even number of consecutive inverter (NOT) gates transmits the logic state of the input to the output. This construction can, however, alter the signal due to degradation effects.\cite{CAV10} In principle, cascaded buffer logic gates (executing the identity operation) can also be used to build a delay line, but they introduce larger degradation.\cite{BEL00b}

Dynamical nodes are built by tailoring the autonomous logic circuit. For example a unidirectional ring of an odd number of autonomous inverter gates---a ring oscillator\cite{HAJ98}---generates periodic square-wave oscillations useful for clock generation\cite{ALV95,SUN01} and for physical random number generation by exploiting the inherent jitter.\cite{HAJ98,SUN07a} It is also possible to assemble logic gates executing other Boolean operations to achieve more complex dynamics such as chaos\cite{ZHA09a} and type-II excitability.\cite{ROS12}

\subsection{Hardware Description Language for an Autonomous Boolean Network}
In the following, we describe typical Verilog code to demonstrate the flexibility in realizing physical design with logic gates on an FPGA and creating networks. An ABN with periodic dynamics is introduced in the next section; its Verilog code reads
\DefineVerbatimEnvironment{code}{Verbatim}{fontsize=\small}
\DefineVerbatimEnvironment{example}{Verbatim}{fontsize=\small}
\newcommand{\mytilde}{$\sim$}
\newcommand{\myvdots}{$\vdots$}
\begin{code}[commandchars=\\\{\}]
module my_osc(s_in,s_out);
 wire [20:0] delay /*synthesis keep*/;
 assign delay[0]  = delay[20] | s_in;
 assign delay[1]  = \mytilde delay[0];
   \myvdots      \myvdots       = \myvdots    \myvdots
 assign delay[20] = \mytilde delay[19];
 assign s_out     =   delay[20];
endmodule
\end{code}
This module, called \verb+my_osc+, describes an ABN with a closed unidirectional chain of inverter (NOT) gates and one OR gate. The NOT and OR logic operations are generated with the $\sim$ and the \verb+|+ operators. The ABN has an input and output \verb+s_in+ and \verb+s_out+, respectively.  The directive \verb+/*synthesis keep*/+ (for Altera FPGAs) guarantees that all logic gates involved with the name \verb+delay+ are implemented by the compiler. Some logic gates would be redundant in synchronous operation and would, therefore, be removed by the compiler. To realize an experimental meta network with two of these ABNs, we consider a  module \verb+main+ in which we call two instanciations of the periodic oscillators \verb+osc1+ and \verb+osc2+ as described by the following hardware description
\begin{code}[commandchars=\\\{\}]
module main(out);
 assign out = net1;
 my_osc osc1(net2, net1);
 my_osc osc2(net1, net2);
endmodule
\end{code}
The output \verb+net1+ of the oscillator called \verb+osc1+ is input to oscillator \verb+osc2+. The output \verb+net2+ of \verb+osc2+ is coupled back, realizing a network of two mutually coupled (here without time delays) dynamical nodes. The output port of the FPGA, called \verb+out+, is connected to \verb+net1+; hence, it will output the dynamics of the first network node. By extending the Verilog code for the main module by a few lines, networks of many nodes can be implemented easily. Note that the definitions of the variables are required at the beginning of the code and are omitted here. By compiling this high-level logical hardware description and loading it on the FPGA, we obtain a true physical (not emulated) network.

\section{Periodic Dynamics in Autonomous Boolean Networks}
\label{sec:Boolean_phase_osc}

In this section, we show that periodic oscillators can be realized and coupled to form networks, thereby achieving phase synchronization. We adapt an existing ring oscillator to ensure unidirectional and bidirectional coupling and observe in-phase and anti-phase synchronization of the oscillator depending on the coupling time delay.

\subsection{Periodic Autonomous Boolean Network: Ring Oscillators}
\label{sec:periodic}

\begin{figure}[b!]
\begin{center}
\resizebox{8.5cm}{!}{\includegraphics{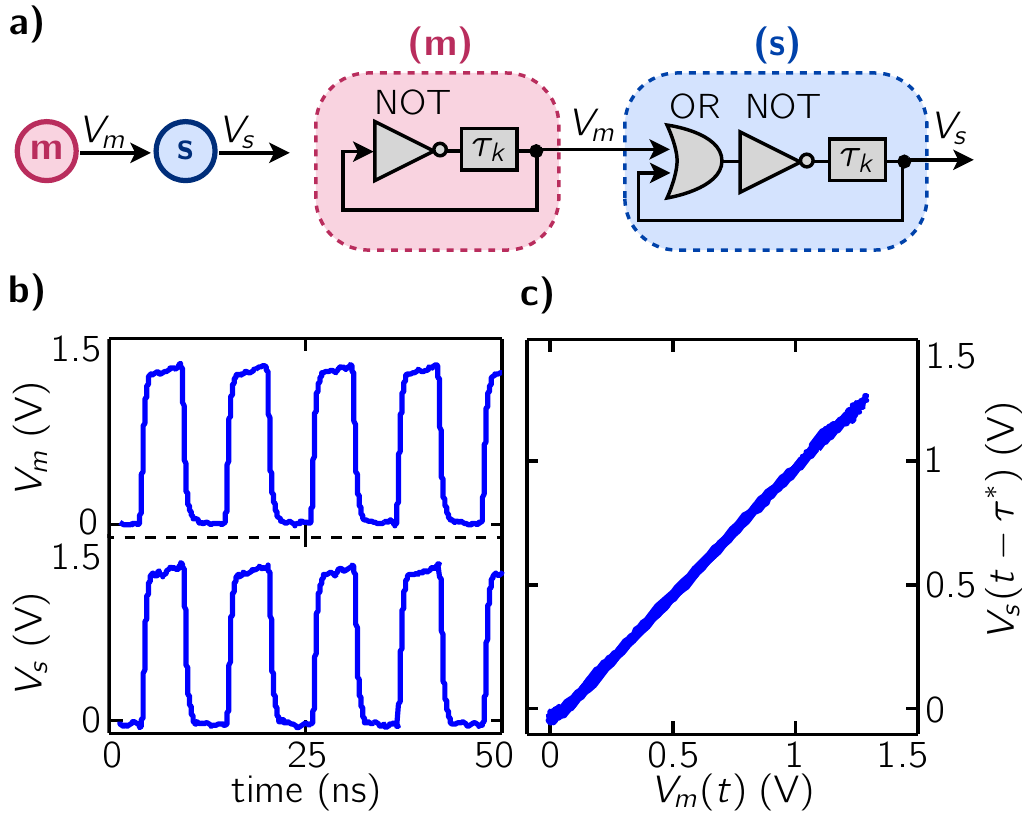}}
\end{center}
\caption{\label{fig:TwoOsc_unidirectional} Experimental demonstration of unidirectional synchronization of Boolean phase oscillators. a) Illustration of the setup with master (m) and slave (s) oscillators. b) Temporal evolutions of the oscillators (m) and (s) showing in-phase square-wave oscillations with period $T_{m,s}=10.9\pm 0.4\;\mathrm{ns}$. c) Evolution in phase plane $(V_m(t),V_s(t-\tau^*))$. The time series are acquired with a high-speed oscilloscope (DSO80804A) with 8 GHz bandwidth and 40 GSa/s sampling rate. }
\end{figure}

A schematic representation of a ring-oscillator design is shown in Fig.~\ref{fig:TwoOsc_unidirectional}a. It comprises one inverter gate subject to time-delayed feedback realized with $n_k$ (even number) inverter logic gates. Not shown are output buffer gates through which the signals pass before they are recorded by an oscilloscope. The design of the ring oscillator prevents the existence of a Boolean fixed point that satisfies simultaneously all inverter logic gates. It displays periodic square-wave oscillations that correspond to a Boolean transition between $V_{L}$ and $V_{H}$ propagating through the ring twice per period.\cite{KAT98} Its fundamental oscillation frequency is $f_k=1/[2(n_k+1)\tau_\mathrm{LG}]$.

%%%%%%%%%%%%%%%%%%%%%%%%%%%%%%%%%%%%%%%%%%%%%%%%%%%%%%%%%%%%%%%%%%%%
%%%%%%%%%%%%%%%%%%%%%%%%%%%%%%%%%%%%%%%%%%%%%%%%%%%%%%%%%%%%%%%%%%%

Usually, ring oscillators are not designed to be coupled. However, a simple modification by the addition of an OR gate allows external Boolean transitions to be injected into the feedback loop.

Using this modified design, we couple two ring oscillators uni-directionally as shown in Fig.~\ref{fig:TwoOsc_unidirectional}a. They are both realized with an identical number of inverter logic gates $n_m+1=n_s+1=21$; with frequencies $f_m =(92.1\pm 0.9)\un{MHz}$ and $f_s =(87.5\pm 1.2)\un{MHz}$, respectively. Their frequencies differ because of the additional OR gate in (s) and heterogeneity in the propagation delay of the logic gates. As a result, the two oscillators are not frequency-locked without coupling. However, when the master oscillator (m) injects its output waveform into the slave oscillator (s), phase- and frequency-locking is achieved with frequency $f_m=f_s=(92.2\pm0.1)\un{MHz}$, as illustrated in Fig.~\ref{fig:TwoOsc_unidirectional}b. Further confirmation of phase synchronization is given in Fig.~\ref{fig:TwoOsc_unidirectional}, where the phase portrait $(V_m(t),V_s(t-\tau^*))$ shows a straight line with slope of one approximately. We measure the quality of synchronization by computing the cross-correlation coefficient between $V_m$ and $V_s$, which is $\rho_{V_mV_s}\approx 0.995$. A skew time $\tau^*\approx 225\;\mathrm{ps}$ is used to compensate for the additional propagation time of the OR gate, the difference in propagation time of the two signals to the output port of the FPGA to the oscilloscope, and for a small propagation delay in the coupling.

The stable phase-locked dynamics corresponds to one Boolean transition propagating in each oscillator with constant relative phase shift. The OR gate used in (s) leads to the creation of a Boolean transition in (s) whenever (m) generates a Boolean transition ($V_{H,L}\to V_{L,H}$) and (s) is in the $V_L$ state. This implies that multiple transitions can potentially propagate in (s) if (m) and (s) are not phase locked. However, the most stable evolution for (s) has a single transition propagating; this results in $V_s(t)$ adjusting to $V_m(t-\tau^*)$. Using an OR gate for the coupling of ring oscillators prevents an accumulation of Boolean transitions in (s): if one of the two inputs is in $V_H$, then a Boolean transition ($V_{H,L}\to V_{L,H}$) in the other input has no influence on the output of the OR logic gate.

Interestingly, such a master-slave architecture realizes a very efficient, yet simple, phase-locked loop (PLL) architecture that does not require a voltage-controlled oscillator, a phase detector, or a complex digital design.\cite{WOL91}

\subsection{Mutual Phase Synchronization of Ring Oscillators}

\begin{figure}[t!]
\begin{center}
\resizebox{8.5cm}{!}{\includegraphics{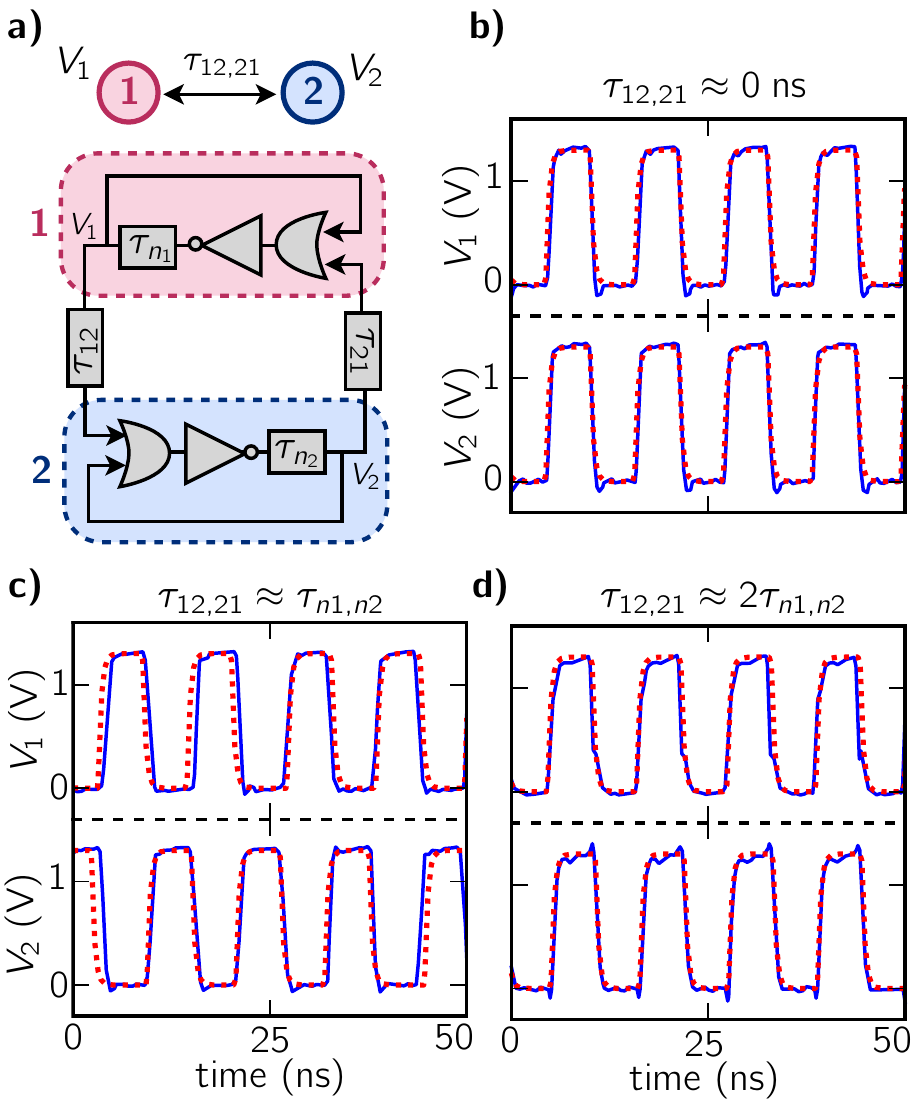}}
\end{center}
\caption{\label{fig:TwoOsc_bidirectional} Demonstration of bidirectional synchronization of two Boolean phase oscillators. a) Boolean implementation of the two oscillators labeled (1) and (2) built with $n_{1}=n_2=21$ inverter gates and coupled by two links with time delays $\tau_{12}\approx\tau_{21}$. b) Temporal evolution of each Boolean oscillator showing in-phase square-wave oscillations with identical period $T_{1}=T_{2}=10.7\pm 0.4\;\mathrm{ns}$ for $\tau_{ij}=0\;\mathrm{ns}$. The time delay is due only to the on-chip wires connecting the two oscillators, and can be neglected ($\tau_{12}\approx\tau_{21}\approx 0$). c)-d) Temporal evolution for the oscillators with $\tau_{12}\approx\tau_{21}\approx\tau_{n_1}\approx\tau_{n_2}$ ($\tau_{12}=6.2\un{ns}$, $\tau_{21}=6.5\un{ns}$) and $\tau_{12}\approx\tau_{21}\approx2\tau_{n_1}\approx2\tau_{n_2}$ ($\tau_{12}=11.7\un{ns}$, $\tau_{21}=11.05\un{ns}$), respectively. The blue solid lines show the experimental time series. The red dotted lines show the dynamics of $x_{buf1}$ and $x_{buf2}$ from numerical simulation of Eqs.~\eqref{eq:DDE_NOT1}-\eqref{eq:DDE_NOT3} with $\tau_1=\tau_2=5.4\un{ns}$ and mutual time delays $\tau_{12}$, $\tau_{21}$ as stated above. The dimensionless quantities $x_{buf1}$ and $x_{buf2}$ are scaled in amplitude and time ($V_{1,2}\rightarrow x_{buf1,2} V_H$ and $t\rightarrow t T_\mathrm{rise}/\ln(2)$, with $V_H=1.3\un{V}$ and $T_\mathrm{rise}=0.26\un{ns}$).
}
\end{figure}

With our modified ring architecture, we can also couple two ring oscillators bidirectionally, as illustrated in Fig.~\ref{fig:TwoOsc_bidirectional}a, with a flexible choice of the coupling time delays $\tau_{12}$ and $\tau_{21}$.

When the coupling time delays are negligible $\tau_{12}\approx\tau_{21}\approx 0\;\mathrm{ns}$, the two oscillators are synchronized in phase, as shown in Fig.~\ref{fig:TwoOsc_bidirectional}b with the frequency of each oscillator being slightly pulled from their respective free-running frequencies $f_1=(81.9\pm0.7)\un{MHz}$ and $f_2=(87.54\pm0.7)\un{MHz}$ to a common frequency $f=(87.7\pm0.7)\un{MHz}$.

The synchronization patterns change when time delays along the links are included. The two ring oscillators displays either in-phase or anti-phase synchronization depending on the coupling time delays $\tau_{12}$ and $\tau_{21}$ with respect to the period of the oscillators $T_{1}\approx2\tau_{n_1}$ and $T_{2}\approx2\tau_{n_2}$. After a series of experiments, we identified that, when $\tau_{12}\approx \tau_{21} \approx p\tau_{n_1}\approx p\tau_{n_2}$ with $p\in\mathbb{N}$ even (odd), the two oscillators are in-(anti-)phase synchronized. To illustrate this, the temporal evolution of each oscillator is shown for $\tau_{12}\approx\tau_{21}\approx \tau_{n_1}\approx \tau_{n_2}$ and $\tau_{12}\approx \tau_{21}\approx 2 \tau_{n_1} \approx 2\tau_{n_2}$ in Fig.~\ref{fig:TwoOsc_bidirectional}c-d, respectively.

Interestingly, our experimental result on mutual synchronization is reminiscent of phase synchronization states predicted theoretically for two coupled Kuramoto oscillators with time-delay feedback loops and links.\cite{DHU08} In their study, however, the periodic oscillator can oscillate without the presence of time-delayed feedback, which is not the case for our Boolean phase oscillator---without the time-delayed feedback, our Boolean oscillator reduces to an OR and a NOT gate with a fixed Boolean state. Similar behavior has been observed numerically for two 
delay-coupled FitzHugh-Nagumo systems, each of which is in the excitable regime, i.e., does not exhibit self-sustained oscillations in the uncoupled case.\cite{SCH08,PAN12}

%%%%%%%%%%%%%%%%%%%%%%%%%%%%%%%%%%%%%%%%%%%%%%%%%%%%%%%%%%%%%%%%%%%%
%% Synchronization in networks
%%%%%%%%%%%%%%%%%%%%%%%%%%%%%%%%%%%%%%%%%%%%%%%%%%%%%%%%%%%%%%%%%%%
In our experiments, the coupling mechanism is the same in both the unidirectional and bidirectional case but differs from typical Kuramoto oscillator, as it can be understood in terms of the exchange of Boolean transitions. The OR gate in each oscillator generates Boolean transitions when the signal of an external Boolean transition is received. Simultaneously, each oscillator maintains a single transition because of the stability associated with a single transition propagating in each ring and the low sensitivity of the OR gate.

\subsection{Model for Ring oscillators}\label{sec:model_ring_osc}
The dynamics of the electronic logic gates are usually modeled from first principles using SPICE models.\cite{VLA94} Here we describe a piecewise-linear switching model of our ABN adapted from previous models of genetic networks\cite{GLA98} and compare the theoretically predicted and experimentally observed dynamics.

The dynamics of the ABN are modeled by considering the continuous output variables of the nodes $x_i(t)$ and the associated Boolean states $X_i(t)$,
\begin{equation}\label{eq:thresh_cond}
X_i(t) = 0 \;\mathrm{ if }\; x_i(t)<x_{th}; \text{  otherwise  } X_i(t) = 1,
\end{equation} 
with the low and high Boolean values 0 and 1 and threshold $x_{th}=0.5$. Specifically, the network in Fig.~\ref{fig:TwoOsc_bidirectional}, which consists of two OR gates with consecutive inverter gates and two delay lines composed of consecutive inverter gates, can be described as two inverted OR (NOR) Boolean functions with delayed feedback. We model the dynamics of this setup, following the formalism introduced by Glass \emph{et al.},\cite{GLA98} and extending it by time delays
\begin{align}\label{eq:DDE_NOT1}
&\frac{\mathrm{d}x_1}{\mathrm{d}t}  = -x_1 + \mathrm{NOR}\left[X_{1}(t-\tau_1),X_{2}(t-\tau_{2,1}-\tau_2)\right],\\
\label{eq:DDE_NOT2}&\frac{\mathrm{d}x_2}{\mathrm{d}t}  = -x_2 + \mathrm{NOR}\left[X_{2}(t-\tau_2),X_{1}(t-\tau_{1,2}-\tau_1)\right],\\
\label{eq:DDE_NOT3}&\frac{\mathrm{d}x_{buf1,2}}{\mathrm{d}t}  = -x_{buf1,2} + X_{1,2}(t),
\end{align}
where NOR$:\left\{0,1\right\}\times\left\{0,1\right\}\rightarrow\left\{0,1\right\}$ denotes the inverted OR operation on the Boolean states. The time delays originate from chains of consecutive inverter gates in the setup ($\tau_1$, $\tau_2$, $\tau_{12}$, $\tau_{21}$). The third equation describes the temporal evolution of two buffer logic gates $x_{buf1}$ and $x_{buf2}$ that perform the Boolean identity operation on $X_{1}(t)$ and $X_{2}(t)$; the buffer gates correspond to output gates on the FPGA.

In Fig.~\ref{fig:TwoOsc_bidirectional}b-d, the dotted red line denotes the solutions obtained from the model for $x_{buf1}$ and $x_{buf2}$ by evolving the analytical solution of the piecewise linear differential equations between the switching of the NOR Boolean function, similar to Ref.~[\onlinecite{GLA98}]. 
Apart from a low level of amplitude noise in the experiment, the dynamics generated by the model agrees well with the experiment. Both display waveforms with an exponential approach to the Boolean states, similar rise times, and similar periodicity of the oscillations. The discrepancy between model and experiment can be quantified via differences in timing of transitions, which is a common measure in autonomous Boolean systems,\cite{ZHA09a} and amounts to average values of $0.20\un{ns}$, $0.94\un{ns}$, and $0.49\un{ns}$ for the waveforms in Fig.~\ref{fig:TwoOsc_bidirectional}b-d, respectively. The error is small in comparison to the oscillation period of $T=10.7\pm0.4\un{ns}$.

\subsection{Discussion}

Above in this section, we show that FPGAs are well suited to realize coupled dynamical systems with periodic dynamics. We assemble the periodic oscillators in simple network motifs and observe phase synchronization. Our experiments display interesting features that are similar to  general theoretical predictions of coupled phase oscillators.\cite{PIK01}

Our approach is scalable to larger network sizes and nodes of higher in-degree. For example, before injecting the input signal into the oscillator, another logic gate with multiple inputs can be used to combine and pre-process multiple input signals from the neighboring nodes.

A limitation of our current approach, however, is the lack of control of the coupling strength. In our design, the coupling is either on or off. In ongoing research, we are developing  an autonomous logic circuit to allow for an adjustable coupling strength so that we can test the various theoretical predictions involving a variation in the coupling strength, such as chimera states\cite{KUR02a,ABR04,OME11} and waves on networks.\cite{SHI04}

\section{Chaotic Dynamics in Autonomous Boolean Networks}

In addition to periodic oscillations, ABNs can display chaos for topologies with multiple loops and multiple-input logic functions.\cite{GHI85,GLA98,ZHA09a,ROS13a} For example, Cavalcante \emph{et al.},\cite{CAV10} showed that chaos emerges in an ABN of two XOR and one XNOR logic gates with links of incommensurate time delays.

In this section, we show that an ABN with a simple topology composed of an XNOR logic gate with three delayed feedback lines also displays chaotic dynamics depending on the time delay of the feedback lines.

\subsection{Realization of a Small ABN with Complex Dynamics}

\begin{figure}[t!]
\resizebox{8.5cm}{!}
{\includegraphics{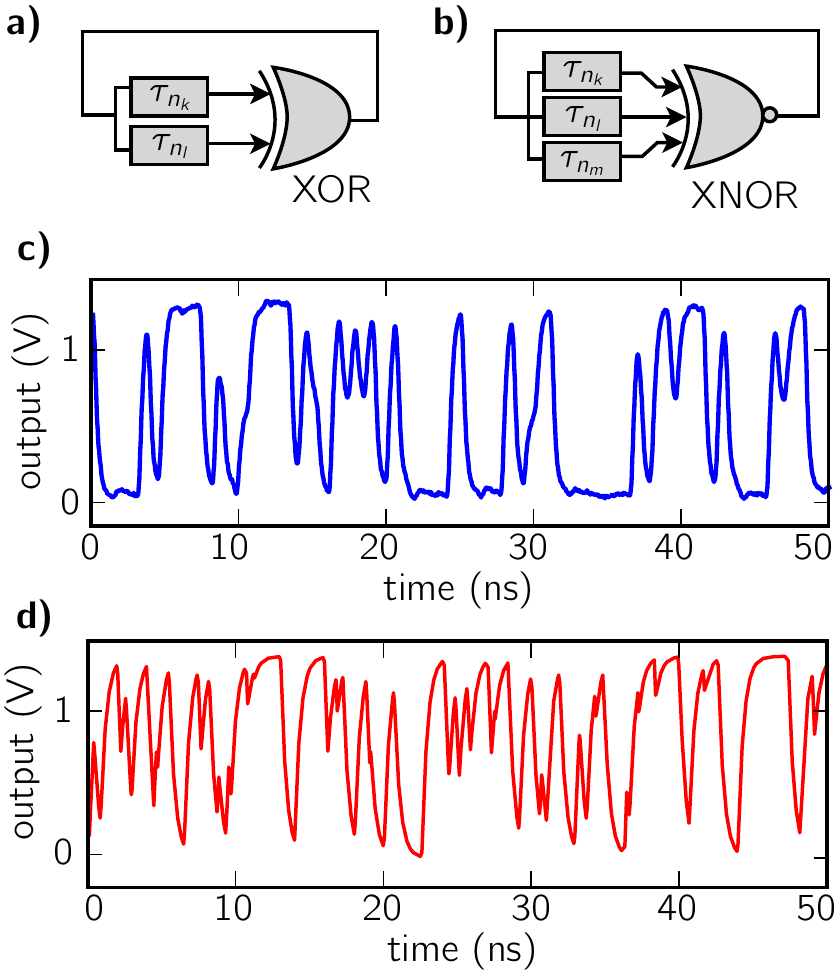}}
\caption{\label{fig:chaotic_xnor}
Experimental demonstration of chaotic dynamics. a) ABN made of one two-input XOR gate with two time delays $\tau_{n_k,n_l}$, as proposed by Ghil \emph{et al.} \cite{GHI85}. b) ABN made of one three-input XNOR gate with three delayed feedback lines. This ABN does not have a Boolean fixed point because $\mathrm{XNOR}(1,1,1)=0$ and $\mathrm{XNOR}(0,0,0)=1$. Not shown are output buffer gates that the signal is routed through prior to the acquisition. c) Chaotic dynamics of the circuit shown in b) for $\tau_{n_k}=(2.8\pm0.1)\un{ns}$,  $\tau_{n_l}=(1.7\pm0.1)\un{ns}$, and  $\tau_{n_m}=(0.56\pm0.02)\un{ns}$ ($n_k=10$, $n_l=6$, $n_m=2$). d) Numerical simulation of Eqs.~\eqref{eq:DDE_XOR}-\eqref{eq:DDE_XOR2} with parameters $\tau_1=3.11\un{ns}$, $\tau_2=1.73\un{ns}$, $\tau_3=0.597\un{ns}$. The dimensionless quantity $x_{buf}$ is scaled in amplitude and time ($V\rightarrow x V_H$ and $t\rightarrow t T_\mathrm{rise}/\ln(2)$, with $V_H=1.3\un{V}$ and $T_\mathrm{rise}=0.26\un{ns}$).}
\end{figure}

Our design of a chaotic dynamical system is motivated by a study of Ghil \emph{et al.},\cite{GHI85} who demonstrated that complex dynamics can emerge in a feedback system comprising one XOR logic gate with two delayed feedback lines of incommensurate time delays $\tau_{n_k,n_l}$ as shown in Fig.~\ref{fig:chaotic_xnor}a. Theoretical analysis  predicts a power-law increase in time of the number of Boolean transition in this circuit. Boolean transitions in the output of the XOR gate are fed back to its input via the two incommensurate delay lines and they trigger new Boolean transitions. In fact, any single change of the input of an XOR logic function leads to a change of the output value, which is called maximum Boolean sensitivity.\cite{KAU04}

This situation, however, cannot occur in our experimental ABN because of the finite bandwidth of logic gates that limits the rate of Boolean transitions.\cite{ZHA09a,CAV10} The low-pass filter effect erases transitions that are too close to each other, a phenomenon called \emph{short-pulse rejection}. Instead, Boolean transitions appear at unpredictable time: A dynamical state referred to as \emph{Boolean chaos}.\cite{ZHA09a}

When realized with electronic logic gates, Ghil's network relaxes to a Boolean fixed state that satisfies the XOR input-output relationship $V_\mathrm{in}=V_\mathrm{out}=V_\mathrm{L}$. This fixed point is always reached after a transient time regardless of the initial conditions and combinations of time delays that we have tested. To observe other dynamics, we need to design a similar network without a Boolean fixed point. For this, we use an XNOR gate (instead of an XOR) and three delayed feedback links, as shown in Fig.~\ref{fig:chaotic_xnor}. The generalization of the XNOR logic operation to more than two inputs corresponds to the inverted parity operation on the Boolean input states.

When implemented on the FPGA, this ABN can display Boolean chaos for a range of values of time delays for each of the three feedback links. Chaotic dynamics is shown in 
Fig.~\ref{fig:chaotic_xnor}c for feedback links with delays of $\tau_{n_k}=(2.8\pm0.1)\un{ns}$, $\tau_{n_l}=(1.7\pm0.1)\un{ns}$, $\tau_{n_m}=(0.56\pm0.02)\un{ns}$ (corresponding to $n_k=10$, $n_l=6$, $n_m=2$ inverter gates, respectively). There is strong evidence of the chaotic nature of the 
waveforms because the mechanism of the generation of Boolean transitions is similar 
to that of an ABN  proven to be chaotic by Zhang \emph{et al.}\cite{ZHA09a} Other evidence is given by the measured fast-decaying and quasi-unstructured autocorrelation function. However, for a
rigorous proof of the chaotic nature, the waveforms should be analyzed more carefully.

%%%%%%%%%%%%%%%%%%%%%%%%%%%%%%%%%%%%%%%%%%%%%%%%%%%%%%%%%%%%%%%%%%%%
%% Figure - Chaotic XOR oscillator
%%%%%%%%%%%%%%%%%%%%%%%%%%%%%%%%%%%%%%%%%%%%%%%%%%%%%%%%%%%%%%%%%%%

\subsection{Model for the Chaotic Oscillator}

Similar to our considerations in Section~\ref{sec:model_ring_osc}, we model the dynamics of the ABN with delay differential equations that switch between two piecewise linear right hand sides 
\begin{align}\label{eq:DDE_XOR}
&\frac{\mathrm{d}x}{\mathrm{d}t}  = -x + \mathrm{XNOR}\left[X(t-\tau_1),X(t-\tau_2),X(t-\tau_3)\right],\\
\label{eq:DDE_XOR2}&\frac{\mathrm{d}x_{buf}}{\mathrm{d}t}  = -x_{buf} + X(t),
\end{align}
where $x(t)$ and $X(t)$ denote the continuous and Boolean state of the XNOR logic gate, respectively, XNOR$:\left\{0,1\right\}\times\left\{0,1\right\}\times\left\{0,1\right\}\rightarrow\left\{0,1\right\}$ denotes the inverted XOR operation on three Boolean states and the time delays originate from the consecutive inverter gates in the setup. The second equation describes, similar to our consideration in Eq.~\eqref{eq:DDE_NOT3}, the temporal evolution of a buffer logic gate $x_{buf}$. 

Figure~\ref{fig:chaotic_xnor}d shows the dynamics obtained from the model for $x_{buf}(t)$, using similar time delays as in the experiment and a threshold voltage of $x_{th}=0.46$ in Eq.~\eqref{eq:thresh_cond}. The numerical dynamics of $x_{buf}(t)$ can be compared to the experimental dynamics in Fig.~\ref{fig:chaotic_xnor}c. Similar features in the two waveforms, such as irregular timing of transitions, can be seen. However, the numerical simulation displays a higher rate of transitions in comparison to the experimental waveform, which returns to the Boolean states for a longer time. In addition, the simulation displays chaos only for narrow ranges of the feedback delays and threshold voltage $x_{th}$, whereas the experiment shows chaotic dynamics consistently for large enough time delays. The differences between experiment and simulations could be due to state-dependency of the feedback delays in the experiment\cite{CAV10} and other non-ideal behaviors not captured by the simplified model.

\subsection{Transition to Chaos}

\begin{figure}[t!]
\resizebox{8.5cm}{!}
{\includegraphics{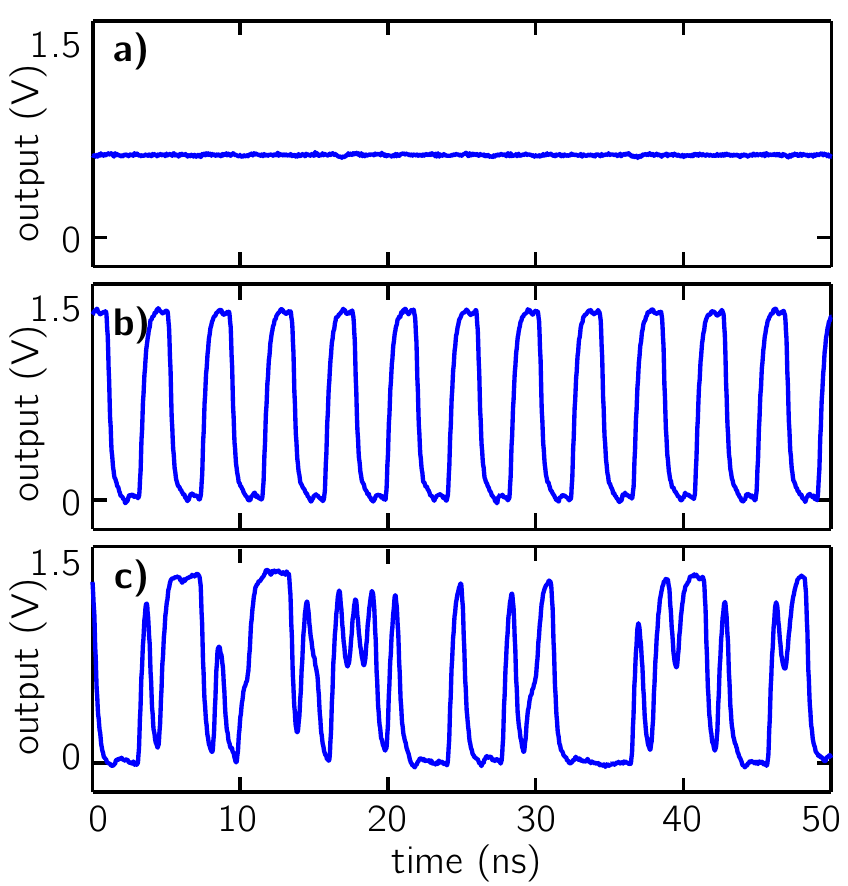}}
\caption{\label{fig:dynamics_XOR_osc}
Dynamics of the oscillator shown in Fig.~\ref{fig:chaotic_xnor} for feedback lines built from fixed numbers of inverter gates $n_l=6$, $n_m=2$, corresponding to $\tau_{n_l}=(1.7\pm0.1)\un{ns}$, $\tau_{n_m}=(0.56\pm0.02)\un{ns}$, and
different numbers $n_k$. a) Fixed point with voltage value between the two Boolean voltages $V_\mathrm{L,H}$ for $n_k=0$ corresponding to $\tau_{n_k}$ of a few picoseconds. b) Oscillatory dynamics with period $T=4.2\pm0.3\;\mathrm{ns}$ for $n_k=2$  $\left[\tau_{n_k}=(1.7\pm0.1)\un{ns}\right]$. c) Chaotic dynamics for $n_k=10$ $\left[\tau_{n_k}=(2.8\pm0.1)\un{ns}\right]$.}
\end{figure}

The network displays chaos only when the time delays of the feedback are sufficiently large. In this section, we keep the two time delays $\tau_{n_l}=\tau_{n_l}=(1.7\pm0.1)\un{ns}$ and $\tau_{n_m}=\tau_{n_m}=(0.56\pm0.02)\un{ns}$ fixed and only vary the value of $\tau_{n_k}$. For short time delay $\tau_{n_k}$, however, regular dynamics is observed. A feedback link of a direct on-chip wire with only a few tens of picoseconds delay leads to a steady-state dynamics. This is because two transitions propagating through that short feedback link in the network will have a frequency that is higher than the cut-off of the low-pass filter of the logic gates. Therefore, the ABN cannot generate Boolean transitions without falling into the previous scenario, thereby only producing a constant voltage at a value that is in between the two Boolean voltage levels, as shown in Fig.~\ref{fig:dynamics_XOR_osc}a. With a short value of the time delay, the threshold value $V_\mathrm{th}$ of the XNOR gate is stabilized.

%%%%%%%%%%%%%%%%%%%%%%%%%%%%%%%%%%%%%%%%%%%%%%%%%%%%%%%%%%%%%%%%%%%%
%% Figure - Dynamics of the XOR oscillator
%%%%%%%%%%%%%%%%%%%%%%%%%%%%%%%%%%%%%%%%%%%%%%%%%%%%%%%%%%%%%%%%%%%

A feedback link of $n_k=2$ inverters, corresponding to a time delay of $\tau_{n_k}=(0.56\pm0.02)\un{ns}$, leads to periodic oscillations, as shown in Fig.~\ref{fig:dynamics_XOR_osc}b. For this value of the time delay, the longest feedback loop is large enough to allow for a single Boolean transition to propagate, while additional transitions cannot be generated by the XNOR gate.

For larger values of $n_k$, the system can display complex dynamics such as quasi-periodicity or periodic oscillations with multiple harmonics. However, the observation of these dynamics depends heavily on the experimental conditions and their existence or the waveform properties may vary significantly when the system is moved to different locations on the FPGA.

Finally, when the number of inverter gates used to realize the feedback time delay reaches or exceeds a threshold value $n_k=10$, the ABN displays Boolean chaos. A measurement of the autocorrelation function calculated from the ABN time series reveals a correlation time of $650 \;\mathrm{ps}$, and the autocorrelation function decays almost to zero for a lag time greater than $100\;\mathrm{ns}$. The threshold value $n_k$ at which the network displays chaos is sensitive to the specific placement of the logic circuit on the FPGA.

\subsection{Discussion}

We demonstrate in this section that ABNs realized on an FPGA can display chaotic dynamics. In agreement with previous studies, our experiments show that the non-ideal behavior of electronic logic gates plays an important role for the dynamics of ABNs.\cite{ZHA09a,CAV10} Short time delays of the feedback loops lead to fixed points and stable periodic dynamics. Longer time delays lead to chaos.

As shown in Section~\ref{sec:Boolean_phase_osc}, it is possible to couple ABNs with periodic oscillatory dynamics to meta-networks and observe synchronization phenomena. However, realizing similar experiments with chaotic ABNs is difficult: Inhomogeneities and inconsistencies in the autonomous mode of operation of logic gates (propagation delays, low-pass filter characteristics, electronic noise) result in significant parameter mismatch when implementing multiple copies of chaotic oscillators on the same FPGA. Consequently, chaos synchronization has not yet been achieved in our experiments. Nevertheless, Boolean chaos in ABNs has several applications. For example, it has already been used for ultra-high-speed random number generation\cite{ROS13a} and is also promising for chaos-based radar applications.\cite{SOB00,LIU04}

\section{Excitable Dynamics in Autonomous Boolean Networks}

In this section, we demonstrate that ABNs can be designed to exhibit excitable dynamics as an artificial neuron and, when connected to a meta-network, they constitute an artificial neural network. We previously used this approach to build small neural-like networks \cite{ROS12,ROS13}, and here we show that we can implement larger networks with random topologies, community structures, and large in-degree of nodes. Building such systems can be useful for understanding large-scale properties of neural systems and for building ultra-fast neuromorphic systems.\cite{IND11}

\subsection{Realization of an ABN with Excitable Dynamics}

Excitability is a property of dynamical systems that generate large excursions in phase space (spikes) in response to small perturbations above a threshold, the stimulus. Such dynamics is often detected for neurons. Another feature of excitable systems is the refractory phase of duration $T_{ref}$, where the excitable system cannot respond to stimuli. We implement excitable systems with these two characteristic features using autonomous logic gates on an FPGA.

\begin{figure}[t!]
\resizebox{8.5cm}{!}{
\includegraphics{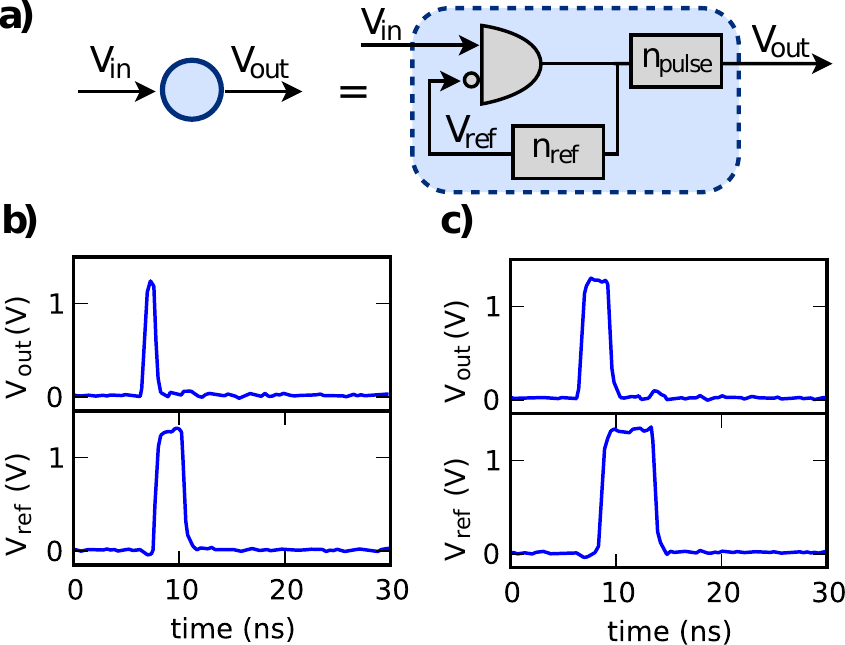}}
\caption{\label{fig:exc_system_setup} Excitable node using an ABN.
a) Scheme of a single excitable node composed of two pulse generators labeled by integers $n_\mathrm{pulse,ref}$ characterizing the pulse width and refractory period: $T_\mathrm{pulse,ref} = n_\mathrm{pulse,ref}\tau_\mathrm{LG}$. b)-c) Temporal evolution of each pulse generator after a single stimulus of the excitable node with b) $n_\mathrm{pulse}=4$ $\left[T_\mathrm{pulse} = (1.12\pm0.04) \;\mathrm{ns}\right]$ and $n_\mathrm{ref}=10$ $\left[T_\mathrm{ref} = (2.8\pm0.1) \;\mathrm{ns}\right]$ and c) $n_\mathrm{pulse}=10$ $\left[T_\mathrm{pulse} = (2.8\pm0.1) \;\mathrm{ns}\right]$ and  $n_\mathrm{ref}=20$ $\left[T_\mathrm{ref} = (5.6\pm0.2)\;\mathrm{ns}\right]$.}
\end{figure}

An excitable system based on an ABN is shown schematically in Fig.~\ref{fig:exc_system_setup}. The excitable node consists of two pulse generators (PGs), which are autonomous logic circuits that generate pulses of constant width in response to a Boolean or continuous voltage that exceeds the threshold voltage of logic gates. The two PGs generate the output voltage of the excitable system $V_\mathrm{out}$ and the refractory voltage $V_\mathrm{ref}$, which indicates the refractory phase. 

We have shown by experiments and numerical simulations\cite{ROS12} of the excitable nodes that these two fundamental properties of neural systems (the pulse generation for above-threshold inputs and the refractory period) leads to basic excitable dynamics that reproduces dynamics of neural networks, such as cluster synchronization, that have been observed previously in complex neuronal models.\cite{ROS13}

The refractory mechanism is implemented with an AND gate that receives inputs from an external input voltage $V_\mathrm{in}$ and $V_\mathrm{ref}$. When the system is (not) in the refractory phase, indicated by $V_\mathrm{ref}=V_{H}$ ($V_\mathrm{ref}=V_{L}$), the AND gate prevents (allows for) an external stimulus to activate the PGs to generate output pulses $V_\mathrm{out}$ and to excite the node. We can adjust the refractory period of the excitable node by changing the width of the pulse in $V_\mathrm{ref}$.\cite{ROS12} We implement an excitable Boolean node on the FPGA and we observe its dynamics in response to a single input stimulus. As shown in Fig.~\ref{fig:exc_system_setup}b and c, the excitable node generates an output and a refractory signal with pulse widths $T_\mathrm{pulse} = (1.12\pm0.06)\;\mathrm{ns}$, $T_\mathrm{ref} = (2.8\pm0.1)\;\mathrm{ns}$ and $T_\mathrm{pulse} = (2.8\pm0.1)\;\mathrm{ns}$, $T_\mathrm{ref} = (5.6\pm0.2)\;\mathrm{ns}$, respectively.

%\textcolor{red}{Pulses are generated and distributed in the network when the continuous input voltages to the excitable nodes cross the gate threshold voltage of $\approx V_H/2$. Consequently, the continuous voltages in ABN that can take values between the Boolean states allows for the all-or-nothing principle of excitablity. The pulse-based signal transmission in ABN of excitable nodes is different than in ABNs with oscillatory and chaotic dynamics, where the signal transmission is driven by the times of Boolean transitions.}

%%%%%%%%%%%%%%%%%%%%%%%%%%%%%%%%%%%%%%%%%%%%%%%%%%%%%%%%%%%%%%%%%%%%
%% Figure - excitable system setup
%%%%%%%%%%%%%%%%%%%%%%%%%%%%%%%%%%%%%%%%%%%%%%%%%%%%%%%%%%%%%%%%%%%

Our excitable node has only a single input $V_\mathrm{in}$. However, in biological neural networks, the in-degree of nodes can be much higher than unity. Therefore, we add another autonomous gate that integrates and combines various incoming signals into a single stimulus. In the literature, such a pre-processing unit is referred to as the synapse of the artificial neuron.\cite{IND11} Here, we use an OR gate---similar to our experiments in Section~\ref{sec:Boolean_phase_osc}. %\textcolor{red}{In addition to the special case of an OR-gate, our experimental platform allows to construct the synapse with arbitrary Boolean functions that can integrate inhibitory and excitatory connections and a Boolean coupling strength. For the latter, the Boolean function is defined such that multiple above-threshold input voltages are required for a high output voltage $V_H$.}

\subsection{Dynamics of Neural-Like Networks of Boolean Excitable Nodes}

\begin{figure}[b!]
\resizebox{8.5cm}{!}{
\includegraphics{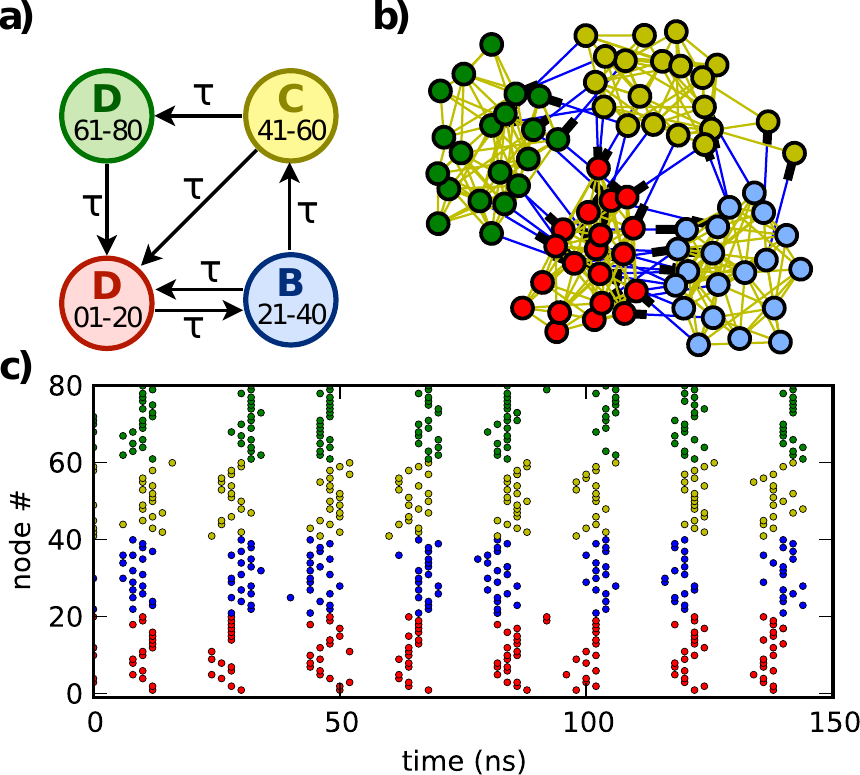}}
\caption{\label{fig:exc_populations}
(Color online) Experimental demonstration of excitable dynamics.
a) Topology of four coupled populations that involves loops of four, three and two elements. (b) Implemented topology, where nodes of the same (different) populations are connected with directed links with probability $P_\mathrm{intra} = 30\%$ ($P_\mathrm{inter} = 1.5\%$). Nodes within a population are strongly coupled with negligible link delay and nodes of different populations are loosely coupled with significant time delay $\tau$. An initial pulse is sent to one node to perturb the network out of its quiescent state. (c) Raster diagram of the network for $n_\tau = 60$ $\left[\tau = (16.8\pm0.6)\un{ns}\right]$, $n_\mathrm{pulse} = 4$ $\left[T_\mathrm{pulse} = (1.12\pm0.04)\un{ns}\right]$, and (b) $n_\mathrm{ref} = 20$ $\left[T_\mathrm{ref} = (5.6\pm0.2)\un{ns}\right]$.
}
\end{figure}

The flexibility of the FPGA allows us to, for example, duplicate excitable Boolean nodes and assemble them in a network of four distinct neural populations, as illustrated in Fig.~\ref{fig:exc_populations}a,b. Dynamical properties of similar networks of excitable systems with community structure have been investigated theoretically\cite{VIC08,KOP12} because of their relevance in analyzing neural circuits such as the thalamic circuitry embedded in the brain.\cite{JON02,SHI03a}

In our experiment, each population consists of $20$ excitable nodes, totaling $80$ nodes for the entire network. The links within a population are connected with on-chip wires on the FPGA so that the associated link time delays are small. The nodes within a population are randomly connected with  probability $p=0.3$. Between the populations, the links are realized with delay lines as defined in Section II with value $\tau=(16.8\pm0.6)\un{ns}$ and probability of connection  $p=0.015$.

%%%%%%%%%%%%%%%%%%%%%%%%%%%%%%%%%%%%%%%%%%%%%%%%%%%%%%%%%%%%%%%%%%%%
%% Figure - Network of coupled populations
%%%%%%%%%%%%%%%%%%%%%%%%%%%%%%%%%%%%%%%%%%%%%%%%%%%%%%%%%%%%%%%%%%%

The dynamics of our artificial neural network is described theoretically by Kanter \emph{et al.}\cite{KAN11a,VAR12} According to the theory, the network dynamics is given by the network topology of the community structure by the greatest common divisor (GCD) of the sizes of directed loops. In the network topology in Fig.~\ref{fig:exc_populations}, inspired by Fig.~1a in Ref.~[\onlinecite{KAN11a}], there are three directed loops of two, three, and four neural populations, respectively. Therefore, the theory predicts a number of synchronized zero-lag synchronized clusters of $\mathrm{GCD}(2,3,4)=1$, i.e., all the populations are predicted to be synchronized with zero time lag.

The experimental dynamics of the network is reported in the raster diagram of Fig.~\ref{fig:exc_populations}c, where each circle corresponds  to a pulse generated by a node. The dynamics of the 80 nodes is acquired with an integrated measurement system based on a processing unit on the FPGA with a timing resolution of $\sim 2\;\mathrm{ns}$. When a pulse is generated by an artificial neuron, its time is recorded and stored on the FPGA on-chip memory.

We observe that all the artificial neurons of the four populations generate pulse trains with period $\tau\pm\Delta\tau$ with $\Delta\tau=5\un{ns}$. 
The dispersion $\Delta\tau$ in the period of the pulse trains originates from heterogeneities in the values of the link time delays and limited resolution 
of our integrated measurement system. The experimental network is considered to be in a near-zero-lag single synchronization cluster,  
which is consistent with the theoretical predictions. This experimental confirmation suggests that our Boolean excitable nodes can be used generically 
to observe other collective phenomena in artificial neural networks.

\subsection{Discussion}
In this section, we measure synchronized neural activity in a neural network of 80 excitable nodes. We show that ABNs can be used to build large meta-networks of neural excitable dynamics. Various synchronization patterns and more general dynamics are expected for high in-degrees of nodes and 
for a different choice of the synapses than an OR gate. For example, the flexibility of the logic function will allow for implementation of inhibiting connections.

Besides potential insights into neurodynamics, our excitable Boolean node may become invaluable for neuro-inspired computing, such as reservoir computing,\cite{JAE04} especially because the nanosecond time-scale of the dynamics will allow for fast processing rates.

\section{Conclusion}

We demonstrate that an FPGA is a versatile experimental platform to conduct integrated experiments on the dynamics of complex networks. When assembling logic gates in their autonomous mode of operation, one can create autonomous Boolean networks (ABNs) that display rich and complex dynamics such as periodic oscillations, chaos, and excitable dynamics. The ABNs with these dynamics can be further coupled with time-delay links to form autonomous Boolean meta-networks that are used to conduct experiments on collective phenomena.

We propose Boolean analogies of three paradigmatic configurations arising in nonlinear dynamics: 
(i) phase synchronization in simple network motifs of oscillatory systems, (ii) chaotic dynamics, and (iii) synchronization phenomena in networks of excitable systems. These three sets of experiments pave the way towards filling the gap between the theory of dynamic networks and desirable experiments, since our approach allows for the realization of large networks with arbitrary topologies on an FPGA.

Nevertheless, our approach still presents many technical and scientific challenges. For example, the experimental extraction of data from each node is only partially solved for networks of excitable systems using the data acquisition capabilities of FPGAs. The greatest challenge in using FPGAs for network experiments is to find the Boolean analogy for the desired dynamical node and the coupling while satisfying technological constraints imposed by the FPGA platform.

\section*{Acknowledgments}
D.P.R., D.R., and D.J.G. gratefully acknowledge the financial support of the U.S. Army Research Office Grants W911NF-11-1-0451 and W911NF-12-1-0099.
E.S. and D.P.R. acknowledge support by the DFG in the framework of SFB~910. We would like to express our thanks to Leon Glass for insightful discussions on the theoretical model.
\bibliographystyle{apsrev4-1}
%\raggedright
%\bibliography{references}

\end{document}